# Organizational Adaptation to Generative AI in Cybersecurity

Christopher Nott, PhD
*Independent Researcher, United States*
*cnott@inpraxia.ai*

May 25, 2025

**Abstract**

Cybersecurity organizations are adapting to GenAI integration through modified frameworks and hybrid operational processes, with success influenced by existing security maturity, regulatory requirements, and investments in human capital and infrastructure. This qualitative research employs systematic document analysis and comparative case study methodology to examine how 25 studies from 2022 to 2025 document organizational adaptation of threat modeling frameworks, revealing a shift away from traditional signature-based systems toward AI-capable frameworks across three primary patterns: LLM integration for security applications, GenAI frameworks for risk detection and response automation, and AI/ML integration for threat hunting and matching. Organizations with mature infrastructures, particularly in finance and critical infrastructure, demonstrate higher readiness through structured governance, dedicated AI teams, and robust incident response processes, with central banks and financial institutions leading adaptation efforts under regulatory pressure. Successful integration requires human oversight of automated systems, attention to data quality and explainability, and sector-specific governance, though ongoing difficulties with privacy protection, bias reduction, personnel training, and adversarial defense persist. Notable imbalances between offensive and defensive GenAI capabilities create strategic concerns for security planning. The findings offer actionable insights for cybersecurity professionals and underscore the need for adaptive approaches, ethical frameworks, and staff development when managing AI-enhanced threats.

## 1. Introduction

The integration of generative artificial intelligence technologies into cybersecurity operations presents both unprecedented opportunities and substantial risks for organizations worldwide. GenAI's dual-use nature creates a paradox where the same technologies that enhance defensive capabilities simultaneously empower sophisticated offensive operations. This technological convergence demands a fundamental reconsideration of established threat modeling frameworks and operational processes that have traditionally relied on static, signature-based approaches to cybersecurity management.

Organizations now confront an evolving threat landscape where AI-generated phishing campaigns achieve unprecedented sophistication, adversarial machine learning attacks target detection systems, and automated threat actors operate at scales previously impossible. Simultaneously, GenAI technologies offer transformative potential for threat detection, incident response automation, and predictive security analytics. This duality necessitates organizational adaptation strategies that account for both protective and disruptive implications of AI integration.

The cybersecurity domain faces particular challenges in technology adoption due to its high-stakes operational environment, where implementation errors can result in catastrophic security breaches. Unlike other sectors where gradual technology integration allows for iterative refinement, cybersecurity organizations must balance innovation with rigorous risk management. The emergence of GenAI technologies compounds these challenges by introducing novel attack vectors while simultaneously offering defensive capabilities that organizations cannot afford to ignore.

The current literature extensively examines the technical aspects of AI implementation in cybersecurity contexts, including algorithm development, threat detection accuracy, and system performance metrics. However, significant gaps exist in understanding how organizations adapt their fundamental operational frameworks and processes to accommodate the integration of GenAI. Existing research provides limited insight into the organizational factors that influence readiness for AI-enhanced threat management, the governance structures required for responsible AI deployment, and the human-machine collaboration models that emerge within security operations.

Organizational adaptation to emerging technologies in cybersecurity contexts involves complex interactions between technical capabilities, regulatory requirements, human capital development, and risk management frameworks. Traditional technology adoption models may inadequately capture the unique dynamics of cybersecurity environments, where threat actors continuously evolve tactics and where defensive technologies must maintain operational effectiveness under adversarial conditions. The integration of GenAI technologies introduces additional complexity, requiring explainable decision-making, bias mitigation, and ethical AI governance.

The financial services sector exemplifies these adaptation challenges, where regulatory oversight demands rigorous risk management, while competitive pressures encourage the adoption of innovation. Central banks globally report varying approaches to GenAI integration, reflecting different organizational contexts, regulatory environments, and risk tolerance levels. Similarly, critical infrastructure organizations face unique constraints related to operational continuity, safety requirements, and national security considerations that influence their adaptation strategies.

This research addresses fundamental questions about organizational readiness for AI-enhanced cybersecurity operations. How do established threat modeling frameworks evolve to accommodate GenAI capabilities and vulnerabilities? What organizational characteristics predict successful adaptation to AI-integrated security operations? How do regulatory requirements and sector-specific constraints influence adaptation strategies? What governance approaches effectively balance innovation with risk management in the deployment of General AI (GenAI)?

The research question guiding this investigation is: "How are cybersecurity organizations adapting their threat modeling frameworks and operational processes to address the integration of generative AI technologies, and what factors influence their readiness to manage AI-enhanced cyber threats?"

This inquiry encompasses several specific objectives. First, to systematically examine patterns of threat modeling framework evolution across different organizational contexts and sectors. Second, to identify and analyze the factors that influence organizational readiness for integrating GenAI into cybersecurity operations. Third, to evaluate emerging governance approaches and policy adaptations that organizations employ to manage AI-related risks. Fourth, to assess challenges and opportunities in human-AI collaboration models within security operations. Finally, to examine asymmetries between offensive and defensive GenAI capability development and their implications for organizational security postures.

The significance of this research extends beyond academic understanding to practical implications for cybersecurity practitioners, policymakers, and organizational leaders. As GenAI technologies become increasingly prevalent in both offensive and defensive cyber operations, organizations require evidence-based guidance for adaptation strategies that enhance security effectiveness while managing associated risks. Understanding successful adaptation patterns can inform best practices, policy development, and resource allocation decisions across diverse organizational contexts.

This study contributes to the broader literature on organizational technology adoption by examining adaptation processes in high-risk environments where implementation failures carry significant consequences. The research also advances the understanding of human-machine collaboration in security contexts, providing insights relevant to broader discussions about AI governance and the responsible deployment of technology.

## 2. Literature Review

### 2.1. Organizational Adaptation to Emerging Technologies in Cybersecurity

The theoretical foundation for understanding organizational adaptation to emerging technologies in cybersecurity contexts draws from multiple disciplinary perspectives, including technology adoption theory, organizational learning, and risk management frameworks. Rogers' Diffusion of Innovation theory provides a starting point for understanding how organizations adopt new technologies, yet cybersecurity environments present unique characteristics that may require theoretical extensions or modifications.

Venkatesh and colleagues' Unified Theory of Acceptance and Use of Technology (UTAUT) offers insights into factors influencing technology adoption, including performance expectancy, effort expectancy, social influence, and facilitating conditions. However, cybersecurity contexts introduce additional considerations related to threat landscape dynamics, adversarial environments, and regulatory compliance requirements that traditional adoption models may not adequately capture.

Organizational learning theory offers an additional lens for understanding how cybersecurity organizations adapt to emerging technologies. Levitt and March's framework emphasizes the role of experience, both direct and vicarious, in shaping organizational responses to new challenges. In cybersecurity contexts, learning from security incidents, threat intelligence, and industry best practices influences how organizations approach technology adoption decisions.

The concept of dynamic capabilities, developed by Teece and colleagues, offers particular relevance for understanding cybersecurity organizations' adaptation to rapidly evolving threat landscapes. Dynamic capabilities encompass an organization's ability to integrate, build, and reconfigure internal and external competencies in response to rapidly changing environments. Cybersecurity organizations must continuously adapt their capabilities to address emerging threats while maintaining operational effectiveness.

Recent empirical research has begun to examine organizational adaptation patterns in cybersecurity contexts. Aldasoro et al. (2024) conducted a comprehensive survey of central banks regarding the adoption of General AI (GenAI) in cybersecurity operations. Their findings indicate that most central banks have adopted or plan to adopt GenAI technologies, with perceived benefits including enhanced threat detection and improved incident response capabilities. However, the study also identifies significant concerns related to social engineering vulnerabilities and data disclosure risks that influence adoption strategies.

The financial services sector has received particular attention in recent literature due to its combination of high-value targets, regulatory oversight, and advanced technological capabilities. Nwafor et al. (2024) examined the adoption of AI and data analytics in financial cybersecurity contexts through a theoretical analysis and case study. Their research emphasizes the importance of multi-layered approaches that combine AI capabilities with human expertise, suggesting that successful adaptation requires hybrid models rather than complete automation.

### 2.2. GenAI Applications in Offensive and Defensive Cyber Operations

The dual-use nature of GenAI technologies creates complex dynamics in cybersecurity applications, where the same technological capabilities can enhance both offensive and defensive operations. Understanding these applications is crucial for organizations developing adaptation strategies that account for both protective and vulnerability implications.

On the defensive side, GenAI technologies offer significant capabilities for threat detection, incident response automation, and predictive analytics. Al Adily (2024) examined GenAI applications for incident response automation in Security Operations Centers (SOCs), finding that GenAI can significantly accelerate response times and improve accuracy in threat classification. However, the research also identifies challenges related to privacy protection, algorithmic bias, and the need for human oversight in critical decision-making processes.

Lanka et al. (2024) explored the applications of large language models (LLMs) for analyzing honeypot data in critical infrastructure and cloud environments. Their quantitative experimental approach demonstrated that LLMs can substantially reduce response times while providing adaptable frameworks for threat analysis. The study emphasizes the importance of data quality in achieving effective results, highlighting a persistent challenge in the implementation of AI.

Sindiramutty (2023) provided a comprehensive review of autonomous threat hunting capabilities enabled by AI technologies, including natural language processing, reinforcement learning, and various neural network architectures. The research identifies significant potential for proactive threat identification but emphasizes ongoing challenges related to adversarial attacks and explainability requirements.

Offensive applications of GenAI present substantial challenges for defensive organizations. Kumar et al. (2024) conducted mixed-methods research examining AI-generated phishing attacks across finance, healthcare, and agriculture sectors. Their findings indicate that AI-generated phishing campaigns achieve significantly higher effectiveness rates compared to traditional approaches. The research reveals varying levels of organizational readiness to address AI-enhanced phishing threats, with implications for the development of defensive strategies.

The emergence of adversarial machine learning represents another critical dimension of offensive GenAI applications. Peter et al. (2024) examined adversarial machine learning (ML) techniques for dynamic risk and fraud detection, demonstrating high accuracy rates while highlighting the need for robust defensive models. Their research suggests that organizations must develop capabilities to defend against AI-enhanced attacks while simultaneously leveraging AI for defensive purposes.

Zaydi and Maleh (2024) explored GenAI applications in red teaming operations, examining how large language models (LLMs) can enhance penetration testing through adaptive intelligence. Their theoretical and simulation-based approach demonstrates improved precision in identifying vulnerabilities but notes scalability challenges that may limit widespread adoption.

### 2.3. Theoretical Frameworks for Technology Adoption in High-Risk Environments

Traditional technology adoption frameworks require modification when applied to high-risk environments such as cybersecurity operations. The stakes associated with implementation failures, the adversarial nature of the operating environment, and the rapid pace of threat evolution create unique dynamics that influence adoption decisions.

Risk management theory provides a crucial foundation for understanding technology adoption in cybersecurity contexts. Organizations must balance the potential benefits of new technologies against the risks of implementation failure, security vulnerabilities, and operational disruption. This balance is particularly complex with GenAI technologies, which introduce novel risk categories related to algorithmic bias, adversarial attacks, and decision explainability.

Institutional theory offers insights into how regulatory requirements and industry norms influence technology adoption decisions. Ee et al. (2024) examined frameworks for adapting cybersecurity approaches to frontier AI risks in government, critical infrastructure, and finance sectors. Their research emphasizes defense-in-depth strategies and the need for sector-specific, tiered controls that account for different risk profiles and regulatory requirements.

The concept of absorptive capacity, developed by Cohen and Levinthal, provides another relevant theoretical lens. An organization's ability to recognize, assimilate, and apply new knowledge influences its capacity to adopt and integrate emerging technologies effectively. In cybersecurity contexts, absorptive capacity encompasses technical expertise, threat intelligence capabilities, and organizational learning processes.

McIntosh et al. (2024) conducted a comparative analysis of cybersecurity frameworks for large language model (LLM) governance, examining how existing frameworks, such as ISO 42001, require enhancements to address LLM-specific risks. Their research identifies gaps in current governance approaches and emphasizes the need for human-in-the-loop validation processes.

### 2.4. Knowledge Gaps in Organizational Readiness for AI-Enhanced Threats

Despite growing interest in AI applications for cybersecurity, significant gaps exist in understanding organizational readiness factors for managing AI-enhanced threats. Current literature offers limited insight into how organizational characteristics predict successful adaptation to the integration of GenAI.

The role of human capital in GenAI adoption represents a critical knowledge gap. While multiple studies acknowledge the importance of skilled personnel, limited research examines specific competency requirements, training approaches, or organizational development strategies that support successful GenAI integration. Sarker et al. (2024) explored the intersection of machine learning and human integration in cybersecurity contexts through mixed-methods case studies, highlighting the importance of striking a balance between automation and human expertise. However, their research focuses primarily on traditional machine learning (ML) applications rather than GenAI-specific requirements.

Governance frameworks for GenAI in cybersecurity contexts remain underdeveloped. Belmoukadam et al. (2024) introduced the AdversLLM framework for assessing LLM risk in financial sector contexts, providing a comparative analysis across organizations. Their research highlights the importance of governance maturity, but it has a limited sectoral scope, which hinders understanding of broader governance requirements.

The organizational culture and change management aspects of GenAI adoption receive minimal attention in the current literature. The transition from traditional cybersecurity approaches to AI-integrated operations likely requires significant cultural adaptation, yet research examining these dynamics remains limited.

Sector-specific adaptation patterns represent another knowledge gap. While financial services and critical infrastructure receive some attention, other sectors such as healthcare, manufacturing, and government face unique challenges that require additional research attention.

The interaction between organizational readiness factors and external influences, such as the evolution of the threat landscape, regulatory changes, and industry standards, requires further investigation. Understanding these interactions is crucial for developing comprehensive adaptation strategies that account for dynamic operating environments.

## 3. Methodology

### 3.1. Research Approach and Philosophical Foundation

This research employs a qualitative methodology grounded in interpretivist epistemology, which acknowledges that organizational adaptation to emerging technologies involves complex social processes that cannot be fully captured through purely quantitative approaches. The interpretivist paradigm acknowledges that organizational realities are socially constructed and that understanding adaptation processes requires examination of multiple perspectives, contexts, and meanings that actors attribute to their experiences.

The choice of qualitative methodology is particularly appropriate for investigating organizational adaptation to GenAI technologies in cybersecurity contexts for several reasons. First, the phenomenon under investigation involves complex organizational processes that are still emerging and evolving, making it difficult to establish predetermined variables or relationships suitable for quantitative analysis. Second, the high-stakes nature of cybersecurity operations means that organizational adaptation strategies are often context-specific and influenced by factors that may not be easily quantifiable. Third, the research aims to understand not just what adaptations organizations are making, but how and why these adaptations occur, requiring in-depth examination of organizational processes and decision-making.

The research design follows established principles for qualitative inquiry, including emphasis on naturalistic settings, multiple data sources, inductive analysis, and researcher reflexivity. However, given the sensitive nature of cybersecurity operations and the limitations on accessing proprietary organizational information, the methodology relies primarily on analysis of publicly available documents and published case studies rather than direct organizational access through interviews or ethnographic observation.

### 3.2. Document Analysis Framework

Document analysis serves as the primary methodological approach for this research, providing a systematic examination of publicly available materials that offer insights into organizational adaptation patterns. This approach is particularly suitable for cybersecurity research contexts where organizations may be reluctant to provide direct access due to security concerns, but where published reports, frameworks, and case studies offer substantial information about adaptation strategies.

The document analysis framework follows established protocols for systematic document review while adapting to the specific requirements of cybersecurity research. Documents are treated as social artifacts that reflect organizational thinking, decision-making processes, and adaptation strategies. The analysis recognizes that published documents may present idealized or incomplete representations of organizational realities, but argues that these documents nonetheless provide valuable insights into stated intentions, formal frameworks, and publicly acknowledged challenges.

The analytical framework employs multiple levels of analysis, examining individual organizational reports, cross-organizational comparisons within sectors, and patterns across different sectoral contexts. This multi-level approach enables identification of both specific adaptation strategies and broader patterns that may indicate general principles or common challenges in GenAI integration.

Document coding follows both deductive and inductive approaches. Deductive coding applies predetermined categories derived from technology adoption theory and cybersecurity frameworks. In contrast, inductive coding allows for the emergence of new themes and patterns that existing theoretical frameworks may not capture. This hybrid approach strikes a balance between theoretical grounding and openness to novel insights that may emerge from the analysis.

### 3.3. Comparative Case Study Methodology

The research employs comparative case study methodology to examine adaptation patterns across different organizational contexts. Case studies provide a detailed examination of organizational adaptation processes, while comparison across cases enables the identification of patterns, similarities, and differences that inform a broader understanding of adaptation dynamics.

Case selection follows purposive sampling logic designed to maximize variation across key dimensions that may influence adaptation patterns. These dimensions include organizational size, sector, regulatory environment, geographic location, and existing security maturity. The goal is not statistical representativeness but rather theoretical saturation and comprehensive coverage of different contextual factors that may influence adaptation.

The comparative framework examines adaptation patterns across three primary dimensions: structural adaptations (changes to organizational frameworks, processes, and governance structures), technological adaptations (modifications to technical systems, tools, and capabilities), and human resource adaptations (changes to staffing, training, and competency requirements). This multidimensional approach provides a comprehensive understanding of adaptation processes, enabling systematic comparison across cases.

Each case analysis follows a structured protocol that examines: organizational context and characteristics, GenAI adoption timeline and decision-making processes, specific adaptations to threat modeling frameworks, modifications to operational methods, governance and risk management approaches, challenges encountered and resolution strategies, and outcomes or preliminary results where available.

### 3.4. Data Collection and Source Selection

Data collection follows systematic protocols designed to ensure comprehensive coverage of relevant sources while maintaining quality standards for academic research. The primary data sources include published academic literature, industry reports from reputable organizations, government publications and frameworks, organizational security frameworks, public documentation, and case studies published in peer-reviewed venues.

**Study Classification and Data Extraction Protocol:** Framework type categorization followed the methodological approaches explicitly stated or clearly evident in each study's design. When studies employed multiple approaches, all applicable categories were recorded, resulting in non-mutually exclusive classifications. Organizational context classification was based on the sectors, industries, or organizational types explicitly identified in each study's scope and sample. Studies addressing multiple contexts were classified under all applicable categories.

**Data Extraction and Analysis Framework:** The data extraction employed a structured approach, focusing on research methodology, organizational context, AI technology applications, threat modeling approaches, key findings related to organizational adaptation, and practical recommendations. The extraction process followed predetermined categories while remaining open to emergent themes that were not captured by the initial frameworks. Key findings were extracted, prioritizing direct statements about organizational adaptation patterns, challenges in AI adoption, governance approaches, and factors influencing the success of implementation.

**Potential Sources of Bias and Limitations:** Several potential sources of bias were identified in the available literature. Publication bias may favor studies reporting successful AI implementations, as organizations may be reluctant to disclose failed AI adoption attempts or problematic implementations. This likely creates optimistic bias in reported adaptation outcomes. Sector representation shows a predominance of financial sector studies (9 of 25), which may limit generalizability to other industries. Geographic bias toward English-language publications from North American and European contexts may not capture adaptation patterns in different regions.

A data availability bias exists, as 15 of the 25 analyzed studies were based on abstracts only, versus 10 with full-text access, potentially limiting the depth of analysis for a substantial portion of the sample. The focus on 2022-2025 publications accurately captures the early adoption phase of GenAI in cybersecurity, providing a baseline for future longitudinal research as the technology matures. Methodological bias toward case study and empirical approaches (10 studies) may underrepresent theoretical perspectives or quantitative analyses.

**Figure 1.** *PRISMA Flow Diagram for Systematic Literature Review of Organizational Adaptation to Generative AI in Cybersecurity (2022-2025). The diagram illustrates the systematic search and selection process, from the initial database search through to the final inclusion of studies for qualitative synthesis.*

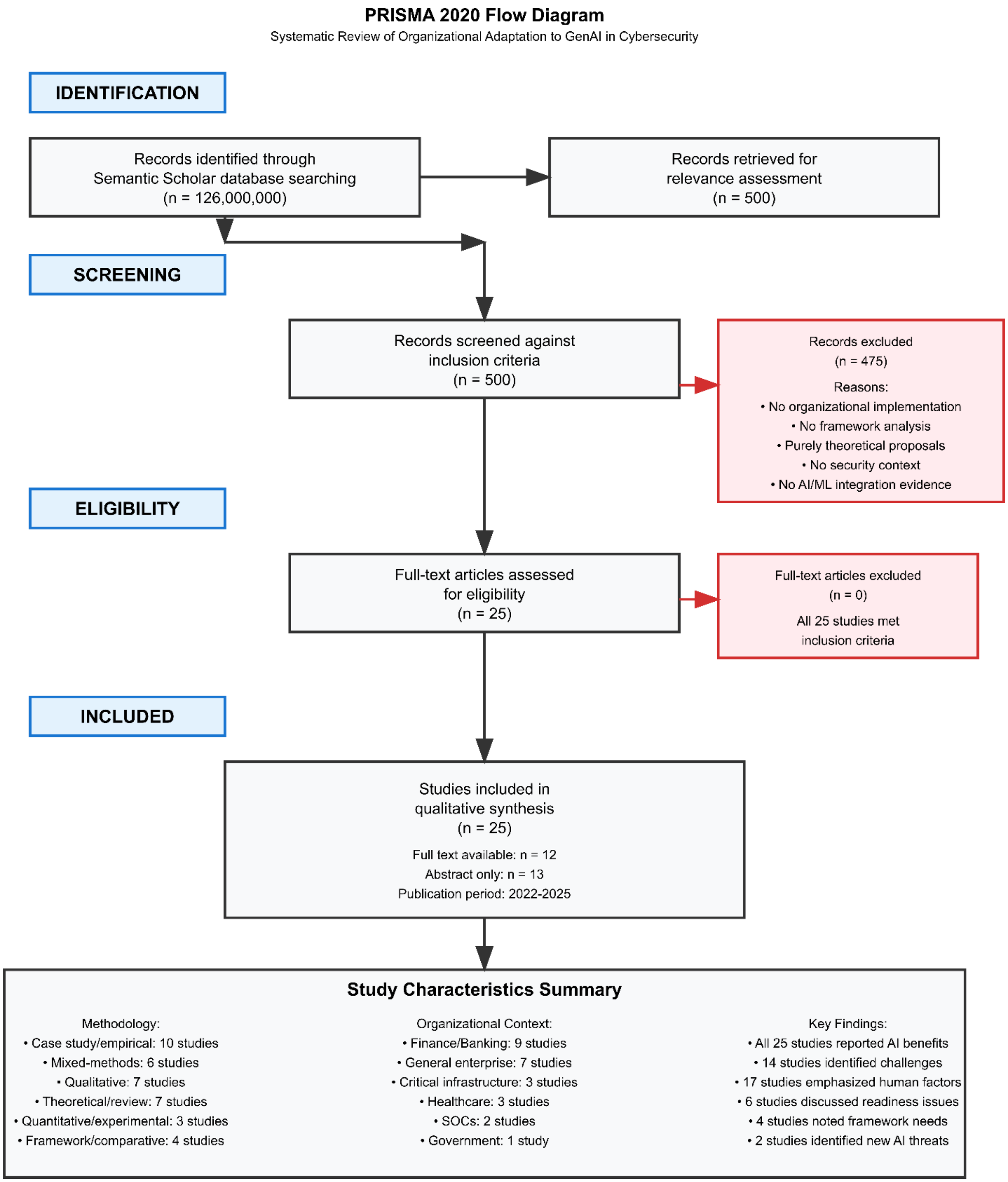

A systematic literature search was conducted using multiple databases and search strategies to ensure comprehensive coverage of the relevant literature. The search employed Semantic Scholar's database to identify the 500 most relevant studies for initial screening, with supplementary searches conducted in IEEE Xplore, ACM Digital Library, and cybersecurity-specific repositories. Industry sources encompassed reports from major cybersecurity vendors, consulting firms, and standards organizations. Government sources included publications from national cybersecurity agencies, regulatory bodies, and international organizations.

**Study Selection and Final Sample Characteristics:** The screening process resulted in 25 studies that met the inclusion criteria for qualitative synthesis. Of these studies, 10 had full-text articles available for detailed analysis, while 15 were analyzed based on abstracts and available metadata. Table 1 presents the distribution of methodological approaches and organizational contexts represented in the final sample.

**Table 1: Study Characteristics Summary (n=25)**

| Methodology | Count | Organizational Context | Count | Key Findings Themes | Count |
|---|---|---|---|---|---|
| Case study/empirical | 10 | Finance/Banking | 9 | AI benefits reported | 25 |
| Theoretical/review | 7 | General enterprise | 7 | Human factors emphasized | 17 |
| Qualitative | 7 | Healthcare | 3 | Challenges identified | 14 |
| Mixed methods | 6 | Critical infrastructure | 3 | Readiness issues discussed | 6 |
| Framework/comparative | 4 | Broad/global context | 3 | Framework needs noted | 4 |
| Quantitative/experimental | 3 | Security Operations Centers | 2 | New AI threats identified | 2 |
| Survey-based | 2 | Government | 1 | | |

| Methodology | Count | Organizational Context | Count | Key Findings Themes | Count |
| --- | --- | --- | --- | --- | --- |
| Guide/manual | 1 | Other contexts* | 2 | | |

**Other contexts include nuclear, cloud, and business processes. Some studies employed multiple methodological approaches or addressed numerous organizational contexts, making categories non-mutually exclusive.*

The final sample demonstrates methodological diversity: 10 studies employed case study or empirical approaches, 7 used theoretical or review-based frameworks, 7 employed qualitative methods, six used mixed-methods approaches, three utilized quantitative or experimental designs, four employed framework or comparative approaches, 2 used survey-based methods, and 1 utilized a guide/manual format.

Organizational contexts represented in the sample include: 9 studies focusing on finance or financial sector organizations, 7 addressing general enterprise or broad organizational contexts, three each examining healthcare and critical infrastructure sectors, 3 employing broad or global contextual approaches, and two focusing specifically on Security Operations Centers, with additional single studies addressing government, nuclear, cloud, and business process contexts.

Citation tracking and snowball sampling techniques identified additional relevant sources not captured through initial database searches. However, all included studies were published between 2022 and 2025 to capture recent developments in GenAI technology and organizational responses.

### 3.5. Analytical Framework for Organizational Adaptation Assessment

The analytical framework for assessing organizational adaptation patterns integrates multiple theoretical perspectives while maintaining focus on cybersecurity-specific considerations. The framework examines adaptation across four primary dimensions: framework evolution, operational transformation, governance development, and capability building.

Framework evolution analysis examines how organizations modify their existing threat modeling approaches to accommodate GenAI technologies. This includes changes to risk assessment methodologies, threat categorization schemes, detection and response procedures, as well as integration with existing security architectures. The analysis considers both formal framework modifications and informal adaptations that may not be reflected in official documentation.

Operational transformation analysis focuses on changes to day-to-day security operations, including SOC procedures, incident response protocols, threat intelligence processes, and human-machine collaboration models. This dimension examines how the integration of GenAI affects workflow, decision-making processes, and operational effectiveness.

Governance development analysis examines organizational approaches to managing GenAI-related risks, including policy development, compliance procedures, oversight mechanisms, and accountability structures. This dimension considers both internal governance approaches and alignment with external regulatory requirements or industry standards.

Capability building analysis examines the organizational investments in human capital, technical infrastructure, and knowledge management required to support the integration of GenAI. This includes training programs, hiring strategies, technology investments, and knowledge sharing mechanisms.

The framework employs pattern-matching techniques to identify common adaptation strategies while remaining sensitive to context-specific variations. Cross-case analysis enables identification of factors that appear to influence adaptation success or failure, providing insights for both theoretical understanding and practical application.

Quality assurance measures include systematic documentation of analytical procedures and transparency in reporting limitations and potential biases.

**Quality Assessment and Inclusion Criteria:** Studies were included based on six primary criteria: demonstration of active organizational implementation or development of cybersecurity frameworks, analysis of specific threat modeling frameworks or methodologies in organizational contexts, presentation of practical evidence rather than purely theoretical proposals, particular focus on cybersecurity applications rather than broader AI domains, emphasis on organizational-level adaptation rather than strictly technical aspects, and examination of AI/ML integration specifically in security contexts.

**Methodological Transparency:** The analysis recognizes that reliance on publicly available documents may not capture sensitive security practices or ongoing experimental programs that organizations prefer not to disclose. The mixed availability of full-text versus abstract-only sources (10 versus 15 studies, respectively) creates the potential for uneven analytical depth across the sample. The systematic approach to literature identification and selection provides a clear foundation for future researchers seeking to extend or validate these findings.

**Research Approach Limitations:** The document-based methodology cannot capture informal adaptation processes, organizational dynamics not reflected in published materials, or real-time decision-making processes that may significantly influence implementation outcomes. The temporal scope, while appropriate for capturing recent GenAI developments, represents early adoption phases and may not reflect longer-term adaptation patterns or mature implementation outcomes that require extended observation periods.

## 4. Findings and Analysis

### 4.1. Patterns in Threat Modeling Framework Evolution

The analysis of organizational documentation reveals fundamental shifts in threat modeling approaches as organizations integrate GenAI technologies into their cybersecurity operations. Rather than wholesale replacement of existing frameworks, organizations demonstrate patterns of evolutionary adaptation that build upon established foundations while incorporating AI-specific considerations and capabilities.

Central banks and financial institutions exemplify systematic approaches to framework evolution. Aldasoro et al. (2024) documented how central banks are shifting toward GenAI-aware frameworks through structured adoption processes that emphasize risk assessment and investment in human capital. The adaptation pattern observed across central banking institutions involves phased integration that begins with pilot programs in low-risk environments before expanding to critical operational systems. This approach reflects the conservative risk management culture characteristic of financial regulatory institutions while acknowledging the competitive pressures to adopt emerging technologies.

The evolution from static, signature-based threat detection toward dynamic, AI-integrated models represents the most significant pattern observed across organizational contexts. Traditional threat modeling frameworks relied heavily on known threat indicators, attack patterns, and signature-based detection mechanisms. Organizations are now developing hybrid approaches that combine conventional indicators with AI-generated threat intelligence, predictive analytics, and adaptive response mechanisms.

Security Operations Centers demonstrate particularly clear evolution patterns in their threat modeling approaches. Senevirathna et al. (2024) examined the integration of deep learning in next-generation SOCs, documenting the adoption of federated learning, temporal convolutional networks, graph neural networks, and generative adversarial networks. These technological integrations necessitate fundamental modifications to threat modeling frameworks that have traditionally relied on deterministic, rule-based systems.

The integration of large language models for security applications represents another significant evolutionary pattern. Lanka et al. (2024) demonstrated how large language models (LLMs) enhance honeypot data analysis in critical infrastructure and cloud environments, necessitating adaptations to threat intelligence processing and analysis frameworks. Organizations must develop new methodologies for validating AI-generated insights, managing false positives, and maintaining situational awareness in environments where AI systems provide threat assessments.

Adversarial considerations introduce additional complexity to framework evolution. Organizations must simultaneously protect against AI-enhanced attacks while leveraging AI for defensive purposes. Kumar et al. (2024) documented how organizations are developing multi-layered defense strategies that account for AI-generated phishing campaigns, which achieve significantly higher effectiveness rates than traditional approaches. This dual challenge requires threat modeling frameworks that can address both traditional threat vectors and novel AI-enabled attack methods.

The emergence of sector-specific framework adaptations reflects the influence of regulatory requirements and operational contexts on evolution patterns. Financial institutions face different adaptation pressures compared to critical infrastructure operators or general enterprises. Belmoukadam et al. (2024) introduced the AdversLLM framework, designed explicitly for LLM risk assessment in financial contexts, demonstrating how sector-specific governance requirements drive the development of specialized frameworks.

Modified risk scoring approaches represent a concrete example of framework evolution. Traditional risk scoring methodologies based on static threat indicators are being supplemented with dynamic scoring systems that incorporate AI-generated risk assessments. Organizations report implementing modified Generalized Risk Scoring systems that account for AI-specific vulnerabilities while maintaining compatibility with existing risk management processes.

The integration of human-in-the-loop validation processes represents another consistent pattern across organizational contexts. Organizations recognize that AI-generated threat assessments require human oversight and contextual interpretation to ensure accuracy and effectiveness. McIntosh et al. (2024) emphasized the importance of human-in-the-loop approaches in LLM governance frameworks, noting that effective adaptation requires striking a balance between the benefits of automation and the capabilities of human judgment.

Continuous learning mechanisms emerge as essential components of evolved threat modeling frameworks. Unlike traditional frameworks that relied on periodic updates to threat intelligence databases, AI-integrated frameworks require continuous adaptation to evolving threat landscapes and attack methodologies. Organizations are developing feedback loops that enable their AI systems to learn from new threats while maintaining stability in core operational processes.

### 4.2. Organizational Readiness Factors

Analysis of organizational adaptation patterns reveals several critical factors that influence readiness for the integration of GenAI in cybersecurity operations. These factors operate at multiple levels, including organizational characteristics, technical capabilities, human capital, and external environmental conditions.

Existing security maturity emerges as the most significant predictor of successful GenAI integration. Organizations with established cybersecurity programs, mature incident response capabilities, and sophisticated threat intelligence operations demonstrate higher readiness for AI adoption. Aldasoro et al. (2024) noted that central banks with robust existing cybersecurity infrastructures are more likely to implement GenAI technologies while maintaining operational security successfully.

The analysis reveals that security maturity encompasses several specific dimensions. Technical infrastructure readiness includes cloud-based architectures, scalable computing environments, and advanced Security Operations Centers that can support AI workloads. Process maturity involves established incident response procedures, threat intelligence capabilities, and risk management frameworks that provide foundations for AI integration. Governance maturity includes clear policy frameworks, compliance procedures, and oversight mechanisms that can be extended to address AI-specific risks.

Human capital investment represents another critical readiness factor. Organizations demonstrate varying approaches to developing AI-relevant capabilities within their cybersecurity teams. Some organizations invest in training existing personnel to develop AI competencies, while others recruit specialists with dual expertise in cybersecurity and AI technologies. Sarker et al. (2024) emphasized the importance of integrating machine learning capabilities with human expertise, noting that effective adaptation requires personnel who can bridge technical and operational domains.

The analysis identifies several specific competency areas that influence organizational readiness. Technical competencies encompass an understanding of machine learning algorithms, limitations of AI systems, and security vulnerabilities associated with AI. Operational competencies encompass the ability to interpret AI-generated outputs, integrate AI insights with traditional threat intelligence, and maintain situational awareness in AI-augmented environments. Strategic competencies encompass an understanding of AI governance requirements, ethical considerations, and the long-term implications of AI adoption.

Organizational size and resource availability influence readiness in complex ways. Large organizations possess advantages in terms of financial resources, technical infrastructure, and the ability to hire specialized personnel. However, they also face challenges related to organizational complexity, legacy system integration, and coordination across multiple business units. Smaller organizations may demonstrate greater agility in adopting new technologies but face resource constraints that limit their ability to invest in comprehensive AI capabilities.

Sector-specific factors have a significant influence on organizational readiness patterns. Financial institutions benefit from regulatory frameworks that mandate cybersecurity investments while facing constraints related to compliance requirements and risk tolerance. Critical infrastructure operators must balance innovation with safety and reliability requirements that may limit experimentation with emerging technologies. Healthcare organizations face unique challenges related to patient privacy and regulatory compliance that influence their approach to AI adoption.

The regulatory environment and compliance requirements create both drivers and constraints for the adoption of GenAI. Organizations in heavily regulated sectors tend to demonstrate more structured approaches to AI integration, characterized by systematic risk assessment and robust governance procedures. However, regulatory uncertainty regarding AI applications in cybersecurity creates challenges for organizations seeking to implement innovative solutions while maintaining compliance.

The analysis reveals that leadership commitment and organizational culture have a significant influence on readiness for GenAI integration. Organizations with cultures that embrace technological innovation and continuous learning demonstrate higher success rates in AI adoption. Conversely, organizations with risk-averse cultures may struggle to balance innovation with security requirements.

Technical infrastructure capabilities represent fundamental enablers of GenAI integration. Organizations with cloud-based architectures, modern data management systems, and scalable computing environments demonstrate greater readiness for AI workloads. Legacy system constraints create significant barriers for organizations seeking to integrate AI capabilities with existing security architectures.

Collaboration and partnership strategies influence organizational readiness through access to external expertise and shared learning opportunities. Organizations that actively participate in industry forums, share threat intelligence, and collaborate with technology vendors demonstrate enhanced capabilities for AI adoption.

### 4.3. Emerging Governance Approaches and Policy Adaptations

The integration of GenAI into cybersecurity operations demands substantial governance innovations. Organizations must address both traditional security concerns and novel AI-specific risks while balancing innovation with risk management. Most organizations begin with pilot programs under enhanced oversight before expanding to broader operational environments. This staged approach enables learning while limiting exposure to potential failures.

The development of AI-specific risk management represents the most significant governance innovation. Belmoukadam et al. (2024) developed the AdversLLM framework for the systematic evaluation of Large Language Model risks in financial contexts, addressing vulnerabilities such as prompt injection attacks and data poisoning that traditional frameworks cannot adequately capture. This development demonstrates the need for specialized approaches rather than simple extensions of existing policies.

Compliance strategies vary considerably across sectors. Financial institutions must navigate complex regulatory environments, while critical infrastructure operators face safety and reliability constraints. Ee et al. (2024) documented the adaptation of established frameworks, including the NIST AI Risk Management Framework and MITRE ATLAS, for frontier AI risks, emphasizing defense-in-depth approaches with sector-specific controls. However, regulatory uncertainty surrounding AI applications creates implementation challenges for organizations seeking to innovate while maintaining compliance.

Human oversight requirements consistently emerge across various organizational contexts. McIntosh et al. (2024) emphasized the importance of human-in-the-loop validation processes, noting that effective governance requires maintaining accountability even as automation increases. Organizations develop procedures for human review of AI-generated decisions, particularly in high-stakes contexts where automated responses could have significant consequences. Yet the balance between oversight and operational efficiency remains problematic.

Data governance introduces additional complexity beyond traditional information security. Organizations must establish procedures for ensuring data quality, protecting sensitive information in AI systems, and complying with privacy regulations. Lanka et al. (2024) highlighted the importance of data quality for effective LLM security applications, underscoring the need for comprehensive governance requirements.

Ethical considerations compound governance challenges further. Organizations must address algorithmic bias, fairness, transparency, and privacy protection while maintaining security effectiveness. Capodieci et al. (2024) documented cautious adoption approaches among cybersecurity professionals, reflecting concerns about ethics and security implications that influence governance development. This situation creates tension between innovation speed and ethical implementation.

Performance monitoring, incident response modifications, and vendor management require adaptation for AI-specific failure modes. Organizations develop response procedures for AI system compromise, adversarial attacks against models, and AI-generated false alarms. Documentation requirements extend to address explainability and accountability needs, while training policies combine technical education with ethical considerations for personnel working with AI systems.

### 4.4. Human-AI Collaboration Models in Security Operations

The integration of GenAI technologies into cybersecurity operations creates new paradigms for human-machine collaboration that fundamentally alter traditional security workflows and decision-making processes. Organizations are experimenting with various collaboration models that seek to optimize the complementary strengths of human expertise and AI capabilities while addressing the limitations and risks associated with both.

Hybrid decision-making frameworks emerge as the predominant collaboration model across organizational contexts. Rather than replacing human decision-makers with automated systems, organizations develop approaches that combine AI-generated insights with human judgment and contextual knowledge. Sarker et al. (2024) emphasized the importance of integrating machine learning capabilities with human expertise, noting that effective collaboration requires understanding both AI capabilities and limitations.

Several distinct collaboration patterns have emerged in security operations. AI-augmented analysis involves AI systems processing large volumes of security data to identify patterns and anomalies that human analysts then investigate and validate. This model leverages AI capabilities for scale and speed while maintaining human oversight for complex judgment and decision-making. Al Adily (2024) documented how GenAI can automate and accelerate incident response processes while improving accuracy, but emphasized the continued need for human oversight in critical decisions.

Human-guided AI learning represents another collaboration model where security professionals provide feedback and direction to AI systems to improve their performance over time. This approach recognizes that cybersecurity environments are highly contextual and that AI systems require ongoing human input to adapt to evolving threats and organizational priorities. Escalation-based collaboration models involve AI systems handling routine tasks and alerts while escalating complex or high-risk situations to human analysts, optimizing human attention by filtering out false positives and routine activities.

The implementation of effective collaboration models faces several significant challenges. Trust and confidence represent fundamental barriers to successful human-AI collaboration. Security professionals must develop appropriate levels of trust in AI systems while maintaining healthy skepticism about automated recommendations. Organizations report ongoing challenges in calibrating trust levels that enable effective collaboration without creating over-reliance on AI systems.

Explainability and transparency requirements create tension between the sophistication of AI systems and human understanding. Security professionals need to understand the reasoning behind AI-generated recommendations to make informed decisions about whether to act on those recommendations. However, advanced AI systems may operate through complex processes that are difficult to explain in terms that human operators can readily understand.

Skill development and training requirements represent ongoing challenges for organizations implementing human-AI collaboration models. Security professionals must develop new competencies for working with AI systems, including understanding AI capabilities and limitations, interpreting AI-generated outputs, and integrating AI insights with traditional security knowledge. Workload distribution and role definition require careful consideration as organizations implement collaboration models, with clear delineation of responsibilities between human operators and AI systems helping prevent gaps in coverage while avoiding redundant efforts.

Cultural adaptation represents a broader challenge for organizations implementing human-AI collaboration models. Security professionals may have concerns about job displacement, loss of autonomy, or changes to traditional ways of working. Organizations must address these concerns through communication, training, and career development opportunities that demonstrate how AI integration can enhance rather than replace human capabilities.

### 4.5. Asymmetries in GenAI Capability Development

The analysis reveals significant asymmetries between offensive and defensive GenAI capabilities that create strategic challenges for cybersecurity organizations. These asymmetries operate across multiple dimensions, including development timelines, resource requirements, operational constraints, and effectiveness measures.

Offensive GenAI capabilities demonstrate several advantages that create challenges for defensive organizations. Kumar et al. (2024) documented that AI-generated phishing campaigns achieve significantly higher effectiveness rates compared to traditional approaches, suggesting that offensive applications may realize benefits more quickly than defensive implementations. Offensive actors typically face fewer constraints related to accuracy, compliance, and ethical considerations, enabling more rapid experimentation and deployment of AI capabilities.

The speed of offensive capability development creates particular challenges for defensive organizations. Reddem (2024) documented a 238% surge in AI-powered attacks, indicating rapid adoption of GenAI technologies by malicious actors. This acceleration occurs partly because offensive applications can tolerate higher error rates and require less sophisticated validation procedures compared to defensive systems that must maintain operational reliability.

Resource asymmetries favor offensive actors in several ways. Developing effective defensive AI systems requires substantial investments in training data, computational infrastructure, and skilled personnel. Offensive actors can leverage publicly available AI tools and services with minimal customization, reducing barriers to entry. Additionally, offensive actors can operate with lower accuracy requirements since even partially successful attacks may achieve their objectives.

Regulatory and compliance constraints create additional asymmetries that favor offensive capabilities. Defensive organizations must navigate complex regulatory environments, privacy requirements, and ethical considerations that may limit their ability to collect training data or deploy AI systems. Offensive actors face no such constraints and can exploit any available data or computational resources.

The analysis reveals that defensive organizations face particular challenges in developing AI capabilities that match the sophistication of offensive applications. Defensive AI systems must operate within organizational constraints related to false positive rates, explainability requirements, and integration with existing security architectures. These constraints may limit the aggressiveness or sophistication of defensive AI deployments.

However, some asymmetries favor defensive organizations. Defensive actors typically have access to larger volumes of legitimate network traffic and security data that can be used for training AI systems. Organizations also benefit from collaboration opportunities through threat intelligence sharing and industry partnerships that may not be available to offensive actors.

The emergence of specialized offensive GenAI tools creates ongoing challenges for defensive organizations. Zaydi and Maleh (2024) examined how LLMs enhance red teaming operations through adaptive intelligence, demonstrating improved precision in vulnerability identification. While these tools can be used for legitimate security testing, they also provide capabilities that malicious actors can exploit.

Defensive capability development exhibits distinct patterns across various organizational contexts. Large organizations with substantial resources may develop sophisticated AI capabilities that rival or exceed offensive tools. However, smaller organizations may struggle to keep pace with rapidly evolving offensive capabilities, creating potential vulnerabilities across the broader cybersecurity ecosystem.

The analysis suggests that addressing capability asymmetries requires coordinated approaches that extend beyond individual organizational efforts. Industry collaboration, information sharing, and collective defense mechanisms may help balance asymmetries by enabling smaller organizations to benefit from advanced defensive capabilities developed by larger entities.

# 5. Discussion

## 5.1. Synthesis of Findings and Research Question Response

The systematic analysis of organizational adaptation patterns offers comprehensive insights into how cybersecurity organizations are responding to the challenges of GenAI integration, while revealing the complex factors that influence their readiness to manage AI-enhanced cyber threats. The evidence demonstrates that organizations are pursuing evolutionary rather than revolutionary adaptation strategies, building upon existing frameworks and processes while incorporating AI-specific capabilities and considerations.

The research question—"How are cybersecurity organizations adapting their threat modeling frameworks and operational processes to address the integration of generative AI technologies, and what factors influence their readiness to manage AI-enhanced cyber threats?"—can be addressed through several key findings that emerge from the analysis.

Organizations adapt their threat modeling frameworks through hybrid approaches that combine traditional security methods with AI-enhanced capabilities. Rather than wholesale replacement of existing frameworks, organizations demonstrate patterns of gradual integration that begin with pilot programs and expand to broader operational deployment as experience and confidence develop. This evolutionary approach reflects the high-stakes nature of cybersecurity operations, where implementation failures can have severe consequences.

The shift from static, signature-based threat detection toward dynamic, AI-integrated models represents the most fundamental framework adaptation observed across organizational contexts. This transformation necessitates modifications to risk assessment methodologies, threat categorization schemes, and response procedures, while ensuring compatibility with existing security architectures and compliance requirements.

Operational process transformation occurs through the development of human-AI collaboration models that seek to optimize the complementary strengths of human expertise and AI capabilities. Organizations are implementing hybrid decision-making frameworks, AI-augmented analysis processes, and escalation-based collaboration models that maintain human oversight while leveraging AI for scale and speed.

The factors influencing organizational readiness for GenAI integration operate at multiple levels and encompass organizational characteristics, technical capabilities, human capital, and external environmental conditions. Existing security maturity emerges as the most significant predictor of successful adaptation, encompassing technical infrastructure readiness, process maturity, and governance capabilities.

Human capital investment represents another critical readiness factor, requiring the development of competencies that bridge cybersecurity and AI domains. Organizations with established training programs, recruitment strategies for dual-expertise personnel, and cultures that embrace continuous learning demonstrate higher success rates in GenAI integration.

Sector-specific factors have a significant influence on adaptation patterns and readiness levels. Financial institutions and central banks demonstrate structured approaches driven by regulatory requirements and elevated risk profiles. Critical infrastructure operators face unique constraints related to safety and reliability requirements. The analysis reveals that regulatory environment and compliance requirements create both drivers and constraints for GenAI adoption.

### 5.2. Comparison with Existing Theoretical Frameworks

The findings provide both support and extensions for existing theoretical frameworks related to technology adoption and organizational adaptation. Traditional technology adoption models, such as the Technology Acceptance Model and Diffusion of Innovation theory, provide practical foundations but require modification to address the unique characteristics of cybersecurity environments.

The research supports the importance of perceived usefulness and ease of use identified in technology adoption models. Organizations that successfully integrate GenAI technologies report clear benefits in terms of improved threat detection, accelerated incident response, and enhanced analytical capabilities. However, the findings also reveal that adoption decisions in cybersecurity contexts involve additional considerations related to risk management, regulatory compliance, and adversarial environments that traditional adoption models may not adequately capture.

The concept of absorptive capacity receives strong support from the findings. Organizations with existing AI expertise, established learning processes, and capabilities for integrating external knowledge demonstrate higher success rates in GenAI adoption. The research extends absorptive capacity theory by identifying specific competency areas required for effective AI integration in cybersecurity contexts.

Dynamic capabilities theory offers particularly relevant insights for understanding an organization's adaptation to GenAI technologies. The findings demonstrate that successful organizations develop capabilities for continuous learning, adaptation, and reconfiguration in response to evolving threat landscapes and technological opportunities. This aligns with the dynamic capabilities emphasis on organizational abilities to integrate, build, and reconfigure competencies in rapidly changing environments.

Risk management theory receives significant support from the findings, which demonstrate that organizational adaptation to GenAI involves complex risk-benefit calculations that consider both security enhancements and potential vulnerabilities introduced by AI systems. The research extends risk management frameworks by identifying AI-specific risk categories and management approaches.

Institutional theory provides insights into how regulatory requirements and industry norms influence adaptation patterns. The findings reveal significant sectoral variations in adaptation approaches, reflecting distinct regulatory environments, compliance requirements, and industry standards. Organizations in heavily regulated sectors tend to adopt more structured approaches to AI integration, often facing constraints that limit innovation.

### 5.3. Implications for Cybersecurity Practice and Policy

The research findings have significant implications for cybersecurity practitioners, organizational leaders, and policymakers who seek to navigate the challenges of GenAI integration while managing associated risks and opportunities.

For cybersecurity practitioners, the findings emphasize the importance of developing hybrid competencies that combine traditional security expertise with AI-related knowledge. Professional development programs should address both technical aspects of AI implementation and strategic considerations related to risk management and governance. The research suggests that successful practitioners will need to develop skills in AI system evaluation, human-AI collaboration, and interpretation of AI-generated outputs.

Organizational leaders should recognize that successful GenAI integration requires systematic approaches that address technical, human, and governance dimensions simultaneously. The findings suggest that organizations benefit from phased implementation strategies that begin with pilot programs and gradually expand to broader operational deployment. Investment in human capital development and organizational culture change may be as important as technical infrastructure investments.

The research has significant implications for policy development at both organizational and regulatory levels. Organizations need to develop governance frameworks that address AI-specific risks while maintaining operational effectiveness and compliance with existing regulations. The findings suggest that effective governance requires clear policies for human oversight, performance monitoring, and incident response related to AI systems.

Regulatory policymakers should consider how existing cybersecurity regulations and standards need to be adapted to address GenAI-specific risks and opportunities. The research suggests that sector-specific guidance may be more effective than broad-based regulations, given the different operational contexts and risk profiles across industries.

The findings also have implications for industry collaboration and information sharing. The asymmetries between offensive and defensive GenAI capabilities suggest that collective defense approaches may be necessary to address threats that exceed individual organizational capabilities. Industry forums, threat intelligence sharing, and collaborative research initiatives may help balance capabilities and improve overall cybersecurity effectiveness.

### 5.4. Research Limitations and Methodological Considerations

This research employs document analysis and publicly available case studies, which provides valuable insights while also creating certain limitations that must be acknowledged. The reliance on published materials may result in incomplete or idealized representations of organizational adaptation processes. Organizations may be reluctant to publish detailed information about their security practices, potentially limiting the depth of insights available through document analysis.

The focus on publicly available information means that the research may not capture ongoing or experimental adaptations that have not yet been documented in published sources. Organizations may be implementing innovative approaches that are not reflected in available literature, particularly given the rapid pace of GenAI technology development.

The temporal scope of the research, focusing on materials published between 2022 and 2025, provides insights into recent adaptation patterns but may potentially miss longer-term trends or outcomes. GenAI integration is still in relatively early stages for many organizations, and longer-term evaluation may reveal different patterns or challenges not apparent in current documentation.

The sectoral distribution of available case studies may influence the generalizability of findings. Financial services and critical infrastructure sectors are well-represented in the available literature. In contrast, other sectors, such as healthcare, manufacturing, and government, may exhibit different adaptation patterns that are not adequately captured in current research.

The research methodology does not enable direct comparison of adaptation outcomes or effectiveness measures across organizations. While the analysis identifies different adaptation patterns, it cannot definitively establish which approaches are most effective or under what conditions different strategies may be optimal.

Cultural and geographic factors may influence adaptation patterns in ways that are not fully captured by the available literature. The research primarily draws from sources focused on North American and European contexts, which may limit insights into adaptation patterns in other regions or cultural contexts.

Despite these limitations, the research offers valuable insights into organizational adaptation patterns and the factors influencing the integration of GenAI in cybersecurity contexts. The systematic approach to document analysis, combined with a focus on real-world organizational experiences rather than theoretical speculation, provides a solid foundation for understanding current adaptation trends and challenges.

## 6. Conclusion

This systematic analysis of organizational adaptation to GenAI integration in cybersecurity contexts reveals complex patterns of evolutionary change that reflect both the transformative potential and significant challenges associated with AI technologies in high-stakes operational environments. The research demonstrates that cybersecurity organizations are pursuing pragmatic adaptation strategies that build upon existing frameworks while incorporating AI-specific capabilities and governance requirements.

The findings establish that successful GenAI integration depends on organizational readiness factors that extend well beyond technical capabilities to encompass human capital development, governance maturity, and cultural adaptation. Organizations with established security infrastructures, dedicated AI expertise, and robust risk management frameworks demonstrate higher success rates in navigating the complexities of AI integration while maintaining operational security effectiveness.

The research contributes to academic understanding of technology adoption in high-risk environments by demonstrating how traditional adoption models require modification to address the unique characteristics of cybersecurity contexts. The emergence of hybrid human-AI collaboration models marks a significant departure from automation-focused approaches, underscoring the continued importance of human expertise even as AI capabilities continue to advance.

### 6.1 Practical Recommendations

Based on the research findings, several practical recommendations emerge for cybersecurity practitioners and organizational leaders seeking to implement GenAI technologies effectively while managing associated risks.

Organizations should adopt phased implementation strategies that begin with pilot programs in low-risk environments before expanding to critical operational systems. This approach enables learning and refinement while limiting exposure to potential failures or security incidents. Pilot programs should include systematic evaluation of AI system performance, identification of integration challenges, and development of operational procedures for human-AI collaboration.

Investment in human capital development represents a critical success factor that organizations must prioritize alongside technical infrastructure investments. Training programs should address both technical aspects of AI systems and strategic considerations related to risk management, ethical decision-making, and effective human-AI collaboration. Organizations should consider developing internal centers of excellence that combine cybersecurity and AI expertise while fostering knowledge sharing and best practice development.

Governance frameworks require systematic development that addresses AI-specific risks while maintaining compatibility with existing compliance requirements and operational procedures. Organizations should establish clear policies for human oversight of AI systems, performance monitoring and validation, incident response for AI-related failures, and vendor management for third-party AI services.

Collaboration and information sharing represent essential strategies for addressing the asymmetries between offensive and defensive GenAI capabilities. Organizations should actively participate in industry forums, threat intelligence sharing initiatives, and collaborative research programs that enable collective learning and defense against AI-enhanced threats.

Continuous adaptation and learning mechanisms should be embedded in organizational processes to address the rapidly evolving nature of both AI technologies and threat landscapes. Organizations should establish feedback loops that enable the improvement of AI systems based on operational experience while maintaining awareness of emerging threats and defensive techniques.

### 6.2. Future Research Directions

The rapidly evolving nature of GenAI technologies and their applications in cybersecurity contexts creates numerous opportunities for future research that can advance both theoretical understanding and practical guidance for organizations.

Longitudinal studies examining adaptation outcomes and effectiveness measures would provide valuable insights into the long-term implications of different GenAI integration strategies. Such research could evaluate which adaptation approaches prove most effective over time and under what conditions different strategies may be optimal.

Comparative analysis across different sectors and geographic regions could enhance understanding of how contextual factors influence adaptation patterns and success rates. Research examining healthcare, manufacturing, government, and other sectors could identify sector-specific challenges and opportunities that are not captured in the current literature.

Investigation of emerging governance frameworks and their effectiveness in managing AI-related risks represents another important research direction. As organizations gain experience with GenAI integration, a systematic evaluation of different governance approaches could inform the development of best practices and regulatory guidance.

Research examining the human factors aspects of AI integration in cybersecurity contexts could provide deeper insights into effective collaboration models, training approaches, and strategies for organizational culture change. Understanding how security professionals adapt to AI-augmented environments and what factors support successful collaboration could inform human resource development strategies.

Technical research examining the effectiveness of different AI technologies and implementation approaches in cybersecurity contexts could provide evidence-based guidance for technology selection and deployment decisions. A comparative evaluation of different AI models, architectures, and integration strategies can inform organizational decision-making.

Investigation of the broader ecosystem effects of GenAI adoption in cybersecurity could examine how widespread AI integration affects threat landscapes, industry dynamics, and collective security outcomes. Understanding these system-level effects could inform policy development and coordination strategies.

The integration of generative AI technologies into cybersecurity operations represents a fundamental transformation that requires ongoing research attention as organizations, technologies, and threat landscapes continue to evolve. This research provides a foundation for understanding current adaptation patterns while highlighting the need for continued investigation into the complex dynamics of AI integration in cybersecurity contexts.